\documentclass[11pt]{article}

\usepackage[draft]{graphics}
\usepackage{graphicx}          	 
\usepackage{bm}                 
\usepackage{amsmath}             
\usepackage{amssymb}
\usepackage{amsfonts}             
\usepackage{verbatim}           
\usepackage{amsthm}             

\usepackage{mathtools}
\usepackage{relsize}

\usepackage{makeidx}

\theoremstyle{plain}             

\theoremstyle{definition}

\makeatletter
\@addtoreset{equation}{chapter}

\makeatother

\def\eqref#1{(\ref{#1})}

\def\Frac#1#2{\frac
{
 {\raise.6ex
 \hbox{$\displaystyle#1$}}
}
{
 {\lower.6ex
 \hbox{$\displaystyle#2$}}
 }
}

\numberwithin{equation}{section}

\def\bigOxe{\sqcup \kern-2.3mm \sqcap}

\makeatother

\def\Frac#1#2{\frac
{
 {\raise.6ex
 \hbox{$\displaystyle#1$}}
}
{
 {\lower.6ex
 \hbox{$\displaystyle#2$}}
 }
}

\input epsf

\def\CHFs#1#2#3{
{}_1F_1\left({a};{c};{z}\right)
}

\def\tfrac#1#2{{{\lower.6ex
\hbox{$\scriptstyle#1$}}\over 
{\raise.7ex
\hbox{$\scriptstyle#2$}}}}

\def\NN{\mathbb N}             

\def\tfrac#1#2{{{\lower.6ex
\hbox{$\scriptstyle#1$}}\over 
{\raise.7ex
\hbox{$\scriptstyle#2$}}}}

\def\insil#1{}

\begin{document}
 \title{Computation of the incomplete gamma function for negative values of
the argument}

\author{
A. Gil\\
Departamento de Matem\'atica Aplicada y CC. de la Computaci\'on.\\
ETSI Caminos. Universidad de Cantabria. 39005-Santander, Spain.\\
   \and
    D. Ruiz-Antol\'{\i}n\\
     Departamento de Matem\'aticas, Estad\'{\i}stica y 
        Computaci\'on,\\
        Univ. de Cantabria, 39005 Santander, Spain. \\ 
\and
J. Segura\\
        Departamento de Matem\'aticas, Estad\'{\i}stica y 
        Computaci\'on,\\
        Univ. de Cantabria, 39005 Santander, Spain.\\
\and
N.M. Temme\\
  IAA, 1825 BD 25, Alkmaar, The Netherlands\footnote{Former address: Centrum Wiskunde \& Informatica (CWI), 
        Science Park 123, 1098 XG Amsterdam,  The Netherlands}\\
}

\date{\ }

\maketitle
\begin{abstract}
An algorithm for computing the incomplete gamma function 
$\gamma^*(a,z)$ for real values of the parameter $a$ and negative real
values of the argument $z$ is presented. The algorithm combines
the use of series expansions, Poincar\'e-type expansions, uniform asymptotic expansions
and recurrence relations, depending on the parameter region. 
A relative accuracy $\sim 10^{-13}$ in the parameter region $(a,z) \in [-500,\,500] \times [-500,\,0)$
can be obtained when computing the function $\gamma^*(a,z)$ with the Fortran 90 module 
{\bf IncgamNEG} implementing the algorithm. 
\end{abstract}

\section{Introduction}

The incomplete gamma function $\gamma^*(a,z)$ is defined by 

\begin{equation}
\label{intro1}
\gamma^*(a,z)=\Frac{z^{-a}}{\Gamma (a)}\gamma(a,z)=\Frac{1}{\Gamma(a)}
\displaystyle\int_0^1 t^{a-1}e^{-zt} dt,
\end{equation}
where $\gamma(a,z)$ is the lower incomplete gamma function \cite[Eqn.~(8.2.1)]{Paris:2010:IGR}. 

The function $\gamma^*(a,z)$ is real for positive and negative values of $a$ and $z$. 

Incomplete gamma functions appear in a large number of scientific applications. For positive values of 
$z$, they are related to the central gamma and chi-squared distribution functions (positive $a$) and
to exponential integrals (negative $a$). There are numerous application areas for positive $z$, for example, 
\cite{Krishna:2006:HSA, Collins:1989:FSA}.
Algorithms and software are available
for this parameter regions \cite{Gautschi:1979:ACP,Didonato:1986:COT,Gil:2012:IGR}. For negative $z$,
the incomplete gamma functions appear, for instance, in the study of Bose plasmas \cite{Kowalenko:1994:AFT,Kowalenko:1998:TMO}
and in the analysis of the Helmholtz equation  \cite{Moroz:1998:QPG,Linton:2010:LSF}. However, unlike the 
positive $z$ case, software to support this case is very limited. Only recently has an algorithm been constructed
for negative $z$ \cite{thompson:2012:IGF} and this is restricted for half-integer values of $a$.

In this paper, we describe an algorithm for computing the function $\gamma^*(a,z)$ for 
$a$ real and $z<0$. Our algorithm improves the range of computation of \cite{thompson:2012:IGF} by allowing real values of $a$.
 The methods of computation used in our algorithm are:

\begin{description}
 \item{a)} series expansions, recurrence relations, and uniform asymptotic expansions for $a<0$; 
\item{b)}  series expansions and Poincar\'e-type expansions \cite[p. 16]{Olver:1997:ASF} for $a>0$.  
\end{description}

A Fortran 90 module implementing the algorithm is provided. Numerical tests show that the relative accuracy is close
to $10^{-13}$ in the parameter 
region  $(a,z) \in [-500,\,500] \times [-500,\,0)$. This module complements
a previous algorithm for the incomplete gamma function for positive values of the 
parameters \cite{Gil:2012:IGR}. 

\section{Methods of computation}

  We describe the methods of computation used in the algorithm and in the numerical tests. Details 
on the region of application of each method are discussed in Section \ref{scheme}.

\subsection{Recurrence relations}\label{RR}

Recurrence relations 
are useful methods of computation when
initial values are available for starting the recursive process.
Also, recurrence relations can be used for testing the function
values obtained by alternative methods.   
Usually, the direction of application of the recursion can not be chosen arbitrarily,
and the conditioning of the computation of a given solution
fixes the direction. 

The function $\gamma^*(a,z)$ satisfies the following inhomogeneous recursion \cite[Eqn.~(8.8.4)]{Paris:2010:IGR}

\begin{equation}
\label{rr1}
z\gamma^*(a+1,z)=\gamma^*(a,z)-\Frac{e^{-z}}{\Gamma(a+1)}
\end{equation}

When both $a$ and $z$ have negative values, replacing $(a,\,z)$ by $(-a-1,\,-z)$ 
and using the reflection formula $\Gamma(a+1)\Gamma(-a)=-\Frac{\pi}{\sin (\pi a)}$ in  (\ref{rr1}),
we obtain \cite[Eqn.~(4.1)]{Temme:1996:UAI}

\begin{equation}
\label{rr2}
\gamma^*(-a-1,-z)+z\gamma^*(-a,-z)=-\Frac{1}{\pi}\sin (\pi a) e^z \Gamma(a+1).
\end{equation}

 We may also combine two first order recursions of (\ref{rr1}) to obtain the three-term homogeneous 
recurrence relation

\begin{equation}
\label{rr3}
z(a+1)\gamma^*(a+2,z)-(a+1+z)\gamma^*(a+1,z)+\gamma^*(a,z)=0.
\end{equation}

Starting from (\ref{rr2}), we obtain

\begin{equation}
\label{rr4}
\gamma^*(-a-2,-z)+(z+a+1)\gamma^*(-a-1,-z)+z(a+1)\gamma^*(-a,-z)=0.
\end{equation}

An advantage of using the relation in (\ref{rr4}) is that possible accuracy problems in the computation
of the inhomogeneous term in (\ref{rr1}) or (\ref{rr2}) are avoided.

\subsection{Series expansion}\label{sec:ser}

A series expansion for $\gamma^*(a,z)$ is given by \cite[Eqn.~(8.7.1)]{Paris:2010:IGR}

\begin{equation}
\label{ser1}
\gamma^*(a, z)=\Frac{1}{\Gamma(a)} \displaystyle\sum_{k=0}^{\infty} \Frac{(-z)^k}{k! (a+k)}\,.
\end{equation}

As pointed
out in \cite{Bailey:2015:CCO} and discussed later (see Section \ref{sec:test}), this series proves to be very useful computationally.
In this form the series cannot be applied when $a=-n,\,\,n=1,\,2,\,\ldots$ and
 special care needs to be exercised when $a=-n+\epsilon$ and $\epsilon$ is small. In this case, it is convenient to rewrite the series as  

\begin{equation}
\label{ser2}
\gamma^*(-n+\epsilon,z)=z^n\Frac{\Gamma(1+n-\epsilon)}{n!}\Frac{\sin \pi \epsilon}{\pi \epsilon}+
    \Frac{1}{\Gamma(-n+\epsilon)}\displaystyle\sum_{k=0,k\neq n}^{\infty}
  \Frac{(-z)^k}{k!(-n+\epsilon+k)}\,.
\end{equation}

Using (\ref{ser2}) the series can be computed as $\epsilon\rightarrow 0$ and we obtain in the limit the result
\cite[Eqn.~(8.4.12)]{Paris:2010:IGR}

\begin{equation}
\label{ser3}
\gamma^*(-n,z)=z^n.
\end{equation}

\subsection{Uniform asymptotic expansion for $a<0$}\label{sec:uae}

When $a$ and $z$ have large negative values, it is convenient to use the uniform
asymptotic expansion described in \cite{Temme:1996:UAI}, where the error function is
used as main approximant. Replacing $(a,\,z)$ with $(-a,\,-z)$ we have

\begin{equation}
\label{uae1}
\gamma^{*}(-a,-z)=z^a\left\lbrace \cos(\pi a)-\sqrt{\frac{2a}{\pi}}e^{\frac{1}{2}a\eta^2}\sin(\pi a)
 \left[\sqrt{\frac{2}{a}}F\left(\eta\sqrt{\frac{a}{2}}\right)+\frac{1}{a}T_a(\eta)\right]\right\rbrace,
\end{equation}
where $\eta$ is defined by
\begin{equation}
\label{uae2}
\frac{1}{2}\eta^2=\lambda-1-\log(\lambda),\quad \lambda=\frac{z}{a},\quad {\rm sign}(\eta)={\rm sign}(\lambda-1).
\end{equation}
The choice of the sign is based on the similarity of the graphs of the $\eta$-function (a parabola) and of the $\lambda$-function (a convex function for $\lambda>0$, with its zero-minimum at $\lambda=1$, and with the shape of a parabola). 

As commented in \cite{Temme:1996:UAI}, it is also useful to consider the normalized function $\widetilde{\gamma}_a(z)$ 
 defined by the relation

\begin{equation}
\label{test1}
\gamma^*(-a,-z)=z^a \cos (\pi a) + \sin (\pi a) \Gamma(a) e^z \widetilde{\gamma}_a(z),
\end{equation} 
giving

\begin{equation}
\label{test2}
\widetilde{\gamma}_a(z)=-\Frac{a}{\pi\Gamma^*(a)}
\left[\displaystyle\sqrt{\frac{2}{a}} F\left(\eta \displaystyle\sqrt{\frac{2}{a}}\right)
+\frac{1}{a}T_a(\eta) \right] \,.
\end{equation}

Using (\ref{test1}) in the inhomogeneous recursion, (\ref{rr2}), we obtain

\begin{equation}
\label{test3}
-\widetilde{\gamma}_{a+1}(z)+\frac{z}{a}\widetilde{\gamma}_a(z) +\frac{1}{\pi}=0.
\end{equation}

In (\ref{uae1}) and (\ref{test2}),  $F(z)$ is Dawson's integral

$$
F(z)=e^{-z^2}\int_0^z e^{t^2}dt = -\frac{1}{2}i\sqrt{\pi} e^{-z^2} \mbox{erf }iz,
$$
where $\mbox{erf }$ is the error function. 

Dawson's integral can be computed using a continued fraction representation. In our algorithm, we
use the representation given in \cite[Eqn.~(13.1.13b)]{Cuyt:2008:HCF}. This continued
fraction works very well for small and large values of $z$.

The function $T_a(\eta)$ in  (\ref{uae1}) and (\ref{test2}) has an asymptotic expansion in negative powers of $a$

\begin{equation}
\label{uae3}
T_a(\eta)\sim\displaystyle\sum_{n=0}^\infty (-1)^n\frac{C_n(\eta)}{a^n},
\end{equation}
where the coefficients, $C_n(\eta)$, may be obtained starting from the differential equation satisfied by $T_a(\eta)$:

\begin{equation}
\label{uae4}
\Frac{d}{d\eta}T_a(\eta) + a \eta T_a(\eta)=a \left(f(\eta)\Gamma^*(a)-1\right),
\end{equation} 
with $f(\eta)$ and $\Gamma^*(a)$ given by

\begin{equation}
\label{ua5}
f(\eta)=\Frac{\eta}{\lambda -1},\,\,\,\Gamma^*(a)=\displaystyle\sqrt{a/(2\pi)}e^aa^{-a}\Gamma(a).
\end{equation}

Substituting the asymptotic expansion (\ref{uae3}) into (\ref{uae4}) and using the expansion
of the reciprocal gamma function 

\begin{equation}
\Frac{1}{\Gamma^*(a)}\sim \displaystyle\sum_{n=0}^{\infty} \Frac{\gamma_n}{a^n},\,\,a \longrightarrow \infty,
\end{equation}
it is possible to find the
following relations for the coefficients $C_n(\eta)$

 \begin{equation}
\label{uae6}
C_0(\eta)=\Frac{1}{\lambda -1}-\Frac{1}{\eta},\,\,\,
\eta C_{n}(\eta)=\Frac{d}{d\eta}C_{n-1}(\eta)+\gamma_nf(\eta),\,n\ge 1.
\end{equation}

When $|\eta|$ is small ($\lambda \rightarrow 1$) the removable singularities
in the representations of the coefficients $C_n$ can be a source of problems in 
numerical computations. 
In \cite{Temme:1996:UAI} Maclaurin expansions for the coefficients $C_0,\,\ldots, C_6$ were used
to generate the values given in Table 4.1 in that reference. In the
present algorithm we use a different approach. Instead of expanding each coefficient, $C_n(\eta)$, 
we expand the function $T_a(\eta)$ of (\ref{uae3}) in powers of $\eta$:

\begin{equation}
\label{uae7}
T_a(\eta)=\sum_{n=0}^\infty \omega_n \eta^n.
\end{equation}

To compute the coefficients, $\omega_n$, we use the differential equation for $T_a(\eta)$
given in (\ref{uae4}). Substituting the expansion (\ref{uae7})
into (\ref{uae4}) and using the coefficients $d_n$ in the expansion

\begin{equation}\label{uae8}
\frac{\eta}{\lambda-1}=\sum_{n=0}^\infty d_n\eta^n,\quad d_0=1,\quad d_1= -\tfrac13,\quad d_2=\tfrac{1}{12},
\end{equation}
we obtain 
\begin{equation}\label{uae8a}
\omega_1=a\left(\Gamma^*(a)-1\right),
\end{equation}
and, for general $\omega_n$, the recursion relation
\begin{equation}\label{uae9}
\omega_n=-\frac{n+2}{a}\omega_{n+2}+d_{n+1}\Gamma^*(a),\quad n=0,1,2,\ldots.
\end{equation}

 If we write 
\begin{equation}\label{uae10}
\omega_n= \alpha_n \Gamma^*(a),\quad n=0,1,2,\ldots,
\end{equation}
we have the recursion
\begin{equation}\label{uae11}
\alpha_n=-\frac{n+2}{a}\alpha_{n+2}+d_{n+1},\quad n=0,1,2,\ldots.
\end{equation}

Then, we choose a positive integer $N$, put $\alpha_{N+2}=\alpha_{N+1}=0$, and
compute the sequence
\begin{equation}
\label{uae12}
\alpha_{N},\alpha_{N-1},\ldots,\alpha_1,\alpha_0
\end{equation}
from the recurrence relation
(\ref{uae11}).

Because (see \eqref{uae8a} and \eqref{uae10})
\begin{equation}
\label{uae13}
\Frac{1}{\Gamma^*(a)}=1-\Frac{1}{a}\alpha_1,
\end{equation}
we have

\begin{equation}
\label{uae14}
T_a(\eta) \approx \frac{a}{a-\alpha_1}\sum_{n=0}^{N} \alpha_n \eta^n
\end{equation}
as an approximation for $T_a(\eta)$.

\subsection{Poincar\'e-type expansion for $a>0$}\label{sec:uaep}
  
  A Poincar\'e-type expansion that is useful for large $|z|$ and valid for all
$a$ bounded can be obtained using the
  relation of $\gamma^*(a,z)$ to the Kummer function
$M(a,b,z)$,

\begin{equation}  
 \gamma^*(a,z)=\Frac{1}{\Gamma(a+1)}\,M(a,1+a,-z),
   \end{equation}
and the expansion given in \cite[Eqn.~(13.7.1)]{Daalhuis:2010:CHF}.  

The resulting expression is given by 

\begin{equation}
  \label{poincare}  
 \gamma^*(a,-z)\sim \Frac{e^z}{z \Gamma(a)} \displaystyle\sum_{n=0}^\infty
\Frac{(1-a)_n}{z^n}.
   \end{equation}

\subsection{Numerical quadrature}\label{sec:int}

For $a>0$, it is also possible to use
numerical quadrature to compute the function $\gamma^*(a,z)$. 
Starting from (\ref{intro1}) we replace $z$ by $-z$

$$
\gamma^*(a,-z)=\Frac{1}{\Gamma(a)}\int_0^1 y^{a-1}e^{zy}dy,
$$

We can then use a quadrature rule to compute this integral to the desired accuracy. One approach
is to consider a change of variable that transforms this integral into one that may be
computed effectively using the trapezoidal rule. A suitable case for this is when the
integrand decays as a double exponential in the real line (see \cite{Takahasi:1973:DEF}
 and \cite[\S5.4]{Gil:2007:NSF}).

We can obtain such an integral representation by using 
the change of variables $r=\text{log}\left(\frac{y}{1-y}\right)$. Then, 

\begin{equation}
\gamma^*(a,-z)= \Frac{1}{\Gamma(a)}\displaystyle
\int_{-\infty}^{\infty} (1+e^{-r})^{-(a+1)}e^{z(1+e^{-r})^{-1}}e^{-r}dr,
\end{equation}
and writing $r=\text{sinh}(t)$, we arrive at

\begin{equation}
\label{int2}
\gamma^*(a,-z)=\Frac{1}{\Gamma(a)}\int_{-\infty}^{\infty}\phi(t)^{a+1}e^{z\phi(t)}e^{-r(t)}\text{cosh}(t)dt,
\end{equation}
where $\phi(t)=(1+e^{-r(t)})^{-1}$.

The integrand of (\ref{int2}) then has double exponential behaviour as $|t|\rightarrow +\infty$ which is suitable for the
application  of the trapezoidal rule. For numerical use, the integral should conveniently truncated by choosing only
a finite interval of integration.

We note here that this quadrature approach is not used in the final version 
of the algorithm as faster methods are available. However, it does provide a useful method
for testing purposes.

\section{Numerical testing and performance}\label{sec:test}

\begin{figure}
\begin{center}
\epsfxsize=14cm \epsfbox{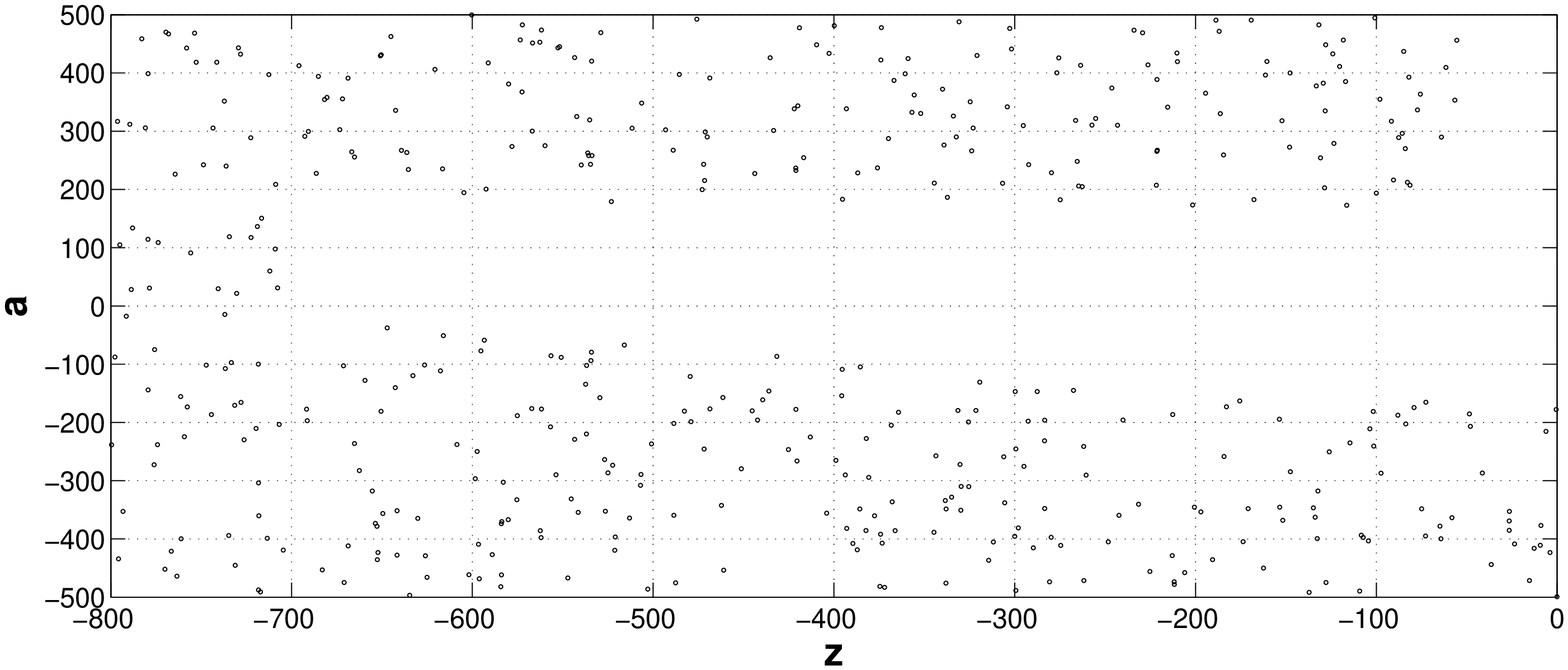}
\caption{Overflow/underflow limitations in double precision arithmetic
in the region $(z,a)\in (-800,\,0) \times (-500,\,500)$ when computing the function
$\gamma^*(a,z)$. The points correspond to values of the variables for which the 
computation either overflows or underflows.
\label{fig1}}
\end{center}
\end{figure}

  For $a<0$, we tested the performance of the uniform asymptotic expansion over a wide range of parameters 
using the normalized gamma function $\widetilde{\gamma}_a(z)$ defined in (\ref{test2}) and the
recurrence relation given in (\ref{test3}). Using $10^8$ random points over the region
of the $(z,a)$-plane $[-1000,\,0) \times [-1000,\,0]$, we obtained an accuracy $\sim 10^{-14}$
in the whole region with the exception of the strips $|a|<4.5$ and $|z|<1.5$.
The range of computation of the $\gamma^*(a,z)$ is more limited due to overflow/underflow problems in 
double precision arithmetic, as can be seen in Figure \ref{fig1}. Function values underflow (overflow) in standard IEEE
double precision arithmetic for large positive (negative) values of $a$. For that reason, we have limited
the rest of the tests to the region of the $(z,a)$-plane $[-500,\,0) \times [-500,\,500]$.

The series expansions of Section \ref{sec:ser} have been tested against a Maple implementation 
using 30 digits accuracy for $a<0$ in the regions
$(z,a) \in [-500,\,0) \times [-5,\,0]$ and $(z,a) \in [-1.5,\,0) \times [-500,\,0]$.  
The maximum relative error  obtained was $\sim 10^{-13}$, although a large number of terms are needed
for computing the series when $|z|$ is large. In this case, a more efficient method of computation is
 to combine the use of recurrence relations and uniform asymptotic expansions. In particular, we compute first the normalized 
gamma function $\widetilde{\gamma}_{\tilde{a}}(z)$ for a value of the parameter $\tilde{a}$ within the
range of validity of the uniform asymptotic expansion and  
then take few steps in the backward direction of the recursion (\ref{test3}). The function $\gamma^*(a,z)$ is finally
computed using (\ref{test1}).  
 
As already mentioned in Section \ref{sec:ser}, we need to be careful in the computation 
when $a$ is close to an integer i.e., $a=-n+\epsilon$ where $\epsilon$ is small. To avoid loss of accuracy both
in the series expansions and when computing the coefficients with the trigonometric functions
in (\ref{uae1}), the input argument, $a$, is defined as a quadruple precision real variable in our implementation.    

 For $a$ positive, testing is made by comparing the available 
methods of computation: series expansions (\ref{ser1}), numerical quadrature (\ref{int2}) and Poincar\'e-type 
expansions (\ref{poincare}). An accuracy 
close to $10^{-14}$ is obtained in the region  $(z,a) \in [-500,\,0) \times [0,\,500]$ using
the series expansion. Numerical quadrature works 
also accurate over the whole region with the exception of $a$-values 
close to zero, where there is some loss of accuracy in the computed function values.
As in the case of $a<0$, the series expansion needs a large number of terms when $|z|$ is large, which makes the use 
of the Poincar\'e-type expansion more efficient for $|z|>50$.

\begin{figure}
\begin{center}
\epsfxsize=14cm \epsfbox{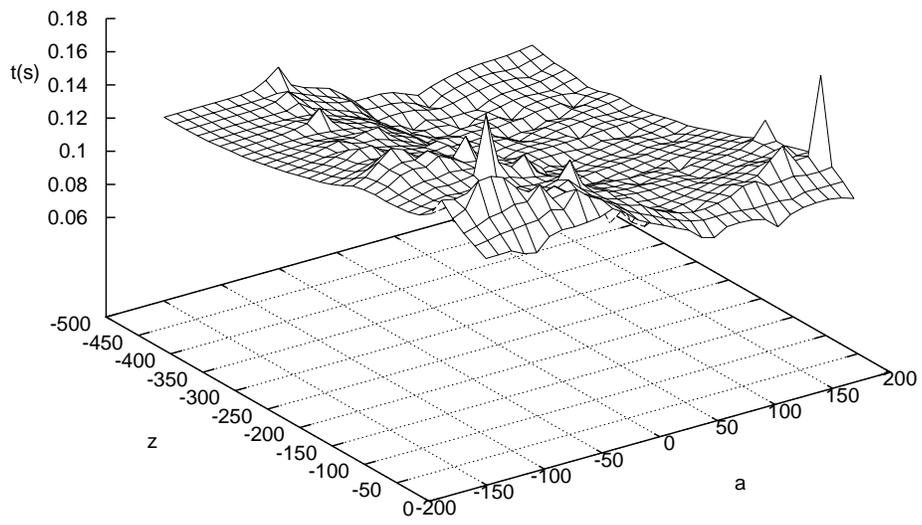}
\caption{CPU time spent by the algorithm as a function of the variables $a$ and $z$. The times shown
correspond to 50000 function evaluations. 
\label{fig2}}
\end{center}
\end{figure}

Figure \ref{fig2} shows the CPU time used by the Fortran version of the algorithm 
in evaluating the function at 50,000 values of $a$ and $z$ on a 2GHz Intel 
Core i5-43100 under Windows 7 Professional. As we can see, the times are quite uniform across the whole range.

\section{Computational scheme}\label{scheme}

From the results obtained in the previous section we may state a stable computational scheme for evaluating the function
$\gamma^*(a,z)$ as follows

\begin{enumerate}
\item{For $a>0$}, 
\begin{description}
\item{If $z<-50$}, compute the function using the Poincar\'e-type expansion~(\ref{poincare}). 
\item{Otherwise,} compute using the series expansion (\ref{ser1}).
\end{description}
\item{For $a<0$},
\begin{description}
\item{If $a=-n,\,n \in \NN$}, use the expression given in (\ref{ser3}).
\item{Otherwise,} 
\begin{description}
\item{If $a>-5$ or $z>-1.5$}, 
\begin{description}
\item{If $z>-100$}, use the series expansion (\ref{ser1})
or the expression (\ref{ser2}) if $a=-n+\epsilon$ and $\epsilon$ is small.
\item{Otherwise,} use the uniform asymptotic expansion (\ref{uae1}) and the recursion relation given in (\ref{test3}).
\end{description}
\item{Otherwise,}  compute the function using the uniform asymptotic expansion (\ref{uae1}). 
\end{description}
\end{description}
\end{enumerate}

\begin{figure}
\begin{center}
\epsfxsize=14cm \epsfbox{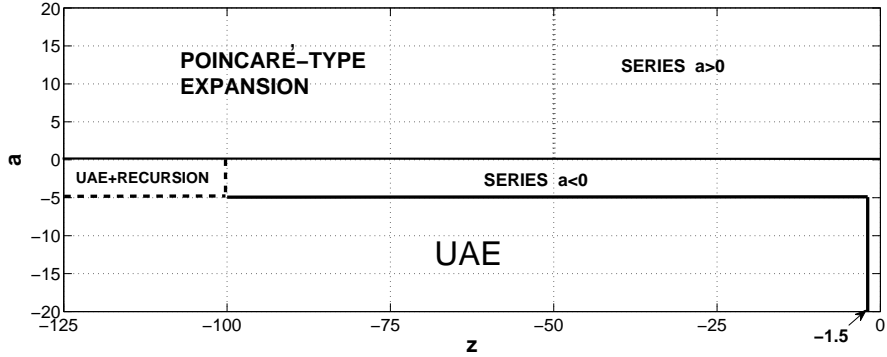}
\caption{Methods for computation of the $\gamma^*(a,z)$ function used in the final algorithm. 
UAE is the uniform asymptotic expansion of Section \ref{sec:uae}. The recursion relation 
is given in Eq. (\ref{test3}).
\label{fig3}}
\end{center}
\end{figure}

 The different methods of computation used in the algorithm, with the exception of
the method for $a=-n,\,n \in \NN$, are shown in Figure \ref{fig3}. 
The domains of computation are established following a
 compromise between efficiency and accuracy: we choose the most accurate method, and where two methods are equally
accurate in a certain parameter region, we choose the fastest.

The resulting algorithm improves the range of computation of the algorithm presented in \cite{thompson:2012:IGF}. Thompson's 
algorithm considers the computation of the lower incomplete
gamma function for negative real values of the argument $z$ and half-integer values of the parameter $a$
using a function $S_n(z)$, $n$ integer and $z>0$, related to the lower incomplete gamma function by
$\gamma(n+1/2,-z)=i(-1)^ne^zz^{n+1}S_n(z)$. The relation of the function $S_n(z)$ to the $\gamma^*(a,z)$ is
then given by $S_n(z)=\Gamma(n+1/2)e^{-z} \gamma^*(n+1/2,-z)$. Precomputed values in Maple 
to initiate analytic continuation are used in Thompson's algorithm which, in the implementation
available in \cite{thompson:2012:IGF}, seems to be restricted to $z$ values in the interval $[0,\,200]$. Our approach
 extends the range of computation to real values 
of the parameter $a$ and larger negative values of the argument $z$, and it does not depend on values precomputed in Maple.

\section{Acknowledgements}

The authors would thank the editors and reviewers for helpful suggestions and comments. 
The authors acknowledge financial support from 
{\emph{Ministerio de Econom\'{\i}a y Competitividad}}, project MTM2012-34787. NMT thanks CWI, Amsterdam, for scientific support.

\bibliographystyle{plain}    
\bibliography{Marcum} 

\end{document}